\documentclass[runningheads]{llncs}
\usepackage[utf8]{inputenc}
\usepackage{latexsym}
\usepackage{amsmath,amssymb,amsfonts}
\usepackage{graphicx}
\usepackage{comment}

\newcommand{\remove}[1]{\relax}

\newcommand{\defeq}{\stackrel{\text{def}}{=}}

\newcommand{\mld}[1]{\ensuremath{{\rm ld}_{\min}(#1)}}
\newcommand{\OPT}{\ensuremath{{\sf OPT}}}
\newcommand{\OPTs}{\ensuremath{\OPT(\sigma)}}
\newcommand{\OPTsa}{\ensuremath{\OPT(\sigma_{\alpha})}}

\newcommand{\dn}[1]{\ensuremath{#1^{-_b}}}
\newcommand{\up}[1]{\ensuremath{#1^{+_b}}}

\def\Alons(#1,#2,#3,#4){\ensuremath{{\sf AACESVW}^{\mbox{\tiny\ensuremath{(#1)}}}_{#3}(#4)}}
\def\Alon(#1,#2,#3){\ensuremath{{\sf AACESVW}^{\mbox{\tiny\ensuremath{(#1)}}}_{#3}}}

\def\DNFa(#1){\ensuremath{\DNF_{#1}}}
\newcommand{\DNF}{\ensuremath{{\sf DNF}}}

\newcommand{\G}{\ensuremath{{\cal G}}}
\newcommand{\Gt}[1]{\ensuremath{\G_{{\hspace*{-0.05em}#1}}}}

\def\Size#1{\left| #1 \right|}

\newcommand{\CASE}[1]{\noindent\textbf{{\bf #1}.} }

\newcommand{\Gttj}{\ensuremath{\G_{\hspace*{-0.05em}t_1 t_2\cdots t_j}}}

\newcommand{\Gtwotwo}{\ensuremath{\G_{\hspace*{-0.05em}22}}}
\newcommand{\Gtwo}{\ensuremath{\G_{\hspace*{-0.05em}2}}}
\newcommand{\GS}{\ensuremath{{\cal G}_{\hspace*{-0.05em}S}}}

\newcommand{\DHk}{\ensuremath{{\sf DH}_k}}
\newcommand{\DHb}{\ensuremath{{\sf DH}_2}}

\newcommand{\DHd}{\ensuremath{{\sf DH}_4}}

\newcommand{\DHllb}{\ensuremath{{\sf DH}^{{\rm lglg}\,n}_2}}
\newcommand{\DHbb}{\ensuremath{{\sf DH}^{b}_2}}

\newcommand{\DWF}{{\sf DWF}}

\newcommand{\aGt}{\ensuremath{\lfloor\hspace*{-0.05em}\alpha\!\cdot\!|\Gtwo|\rfloor}}
\newcommand{\apGt}{\ensuremath{\lfloor\hspace*{-0.05em}\alpha'\!\!\cdot\!|\Gtwo|\rfloor}}
\newcommand{\aLGt}{\ensuremath{\lfloor\hspace*{-0.05em}\alpha_L\!\!\cdot\!|\Gtwo|\rfloor}}

\newcommand{\dGt}{\ensuremath{\lfloor\hspace*{-0.05em}\delta\!\cdot\!|\Gtwo|\rfloor}}

\newcommand{\gta}{\ensuremath{g}}
\newcommand{\nG}{\ensuremath{n_{\hspace*{-0.07em}g}}}
\newcommand{\nB}{\ensuremath{n_{\hspace*{-0.07em}B}}}
\newcommand{\mR}{\ensuremath{m_{\hspace*{-0.07em}R}}}
\newcommand{\mRB}{\ensuremath{m_{\hspace*{-0.08em}R\hspace*{-0.06em}B}}}
\newcommand{\sB}{\ensuremath{s_{\hspace*{-0.07em}B}}}
\newcommand{\xB}{\ensuremath{x_{\hspace*{-0.07em}B}}}
\newcommand{\eB}{\ensuremath{e_{\hspace*{-0.07em}B}}}
\newcommand{\mB}{\ensuremath{m_{\hspace*{-0.07em}B}}}
\newcommand{\mW}{\ensuremath{m_{\hspace*{-0.07em}W}}}

\begin{document}
\title{Improving Online Bin Covering with Little Advice}
%
%

\author{
Andrej Brodnik\inst{1}\orcidID{0000\,0001\,9773\,0664} \and
Bengt J.~Nilsson\inst{2}\orcidID{0000\,0002\,1342\,8618} \and
Gordana Vujović\inst{3, 4}\orcidID{0000\,0001\,5283\,3440}
}

\institute{
University of Ljubljana, Ljubljana, Slovenia.\hspace{0.5em}
\email{andrej.brodnik@fri.uni-lj.si}
\and
Malmö University, Malmö, Sweden.\hspace{0.5em}
\email{bengt.nilsson.TS@mau.se} 
\and
(Student) University of Ljubljana, Ljubljana, Slovenia.\hspace{0.5em}
\and 
(Student) Complexity Science Hub, Vienna, Austria.\hspace{0.5em}
\email{vujovic@csh.ac.at}}

\authorrunning{A.~Brodnik, B.\,J.~Nilsson, G.~Vujović}

\maketitle              
\begin{abstract}
The online bin covering problem is: given an input sequence of items find a placement of the items in the maximum number of bins such that the sum of the items' sizes in each bin is at least~1.  Boyar~{\em et~al}.\@~\cite{boyar2021} present a strategy that with $O(\log \log n)$ bits of advice, where $n$ is the length of the input sequence, achieves a competitive ratio of $8/15\approx0.5333\ldots$. We show that with a strengthened analysis and some minor improvements, the same strategy achieves the significantly improved competitive ratio of~$135/242\approx0.5578\ldots$, still using $O(\log \log n)$ bits of advice.
\end{abstract}

\section{Introduction}\label{sec:intro}

In the bin covering problem, we are given a set of items of 
sizes in the range $]0,1[$ and the goal to cover a maximum number of bins where a bin is covered if the sum of sizes of items placed in it is at least~$1$. It has been shown that the problem is NP-hard~\cite{ASSMANN1984:bincovering,assmann1983problems}.

The bin covering problem has applications in various situations in business and industry, from packing medical relief boxes to emergency areas so that each box contains at least a minimum net weight, to such complex problems as
distribution of data/items to a maximum number of processors/bins, where the processor can run a task if at least some minimal amount of data is available.
%

In the online version, items are delivered successively (one-by-one) and each item has to be packed, either in an existing bin or a new bin, before the next item arrives. 
The first online strategy proposed for the problem is {\em Dual Next Fit\/} (\DNF) and its competitive ratio of $1/2$ is proved by Assmann~{\em et~al}.~\cite{ASSMANN1984:bincovering}. Later Csirik and Totik~\cite{CSIRIK1988163} prove that no pure online strategy can achieve a better competitive ratio.


Boyar~{\em et~al}.~\cite{boyar2021} look at bin covering using extra advice provided by an oracle through an advice tape that the strategy can read. If the input sequence consists of $n$ items, they show that with $o(\log\log n)$ bits of advice, no strategy can have better competitive ratio than $1/2$. In addition, they show that a linear number of bits of advice is necessary to achieve competitive ratio greater than~$15/16$. Finally, they provide a strategy using $O(\log\log n)$ bits of advice and competitive ratio $8/15-O(1/\log n)$.
We thus differentiate between pure strategies and advice-based strategies.

Brodnik~{\em et~al.}~\cite{matcos2022}, show a $2/3$-competitive strategy using $O(\zeta+\log n)$ advice, where $\zeta$ is the number of bits used to encode any item value in~$\sigma$. 
However, this result does not follow the model used by Boyar~{\em et~al}, in that they make assumptions on the encoding of the input items.

\paragraph*{Our Results.} We present an improved analysis of the online strategy by Boyar~{\em et~al}.~\cite{boyar2021} and show that with additional modifications it is possible to achieve a competitive ratio of~$135/242-O(1/\log n)$ using $O(\log\log n)$ bits of advice.

\section{Preliminaries}\label{sec:prel}

The \emph{online bin covering problem\/} we consider is, given an input sequence $\sigma=(v_{1},v_{2},\ldots)$ of $\Size{\sigma}=n$ values $v_{i}$ with
$v_{i} \in\ ]0,1[$, 
find the \emph{maximum} number of unit sized bins that can be covered online with items from the input sequence~$\sigma$.
A bin $B$ is covered if $\sum_{v \in B} v \geq 1$. %

We measure the quality of an online maximization strategy ${\sf A}$ by its \emph{asymptotic competitive ratio}, the maximum bound $R$ such that

\vspace*{-2ex}
\begin{equation}
\Size{{\sf A}(\sigma)} \geq R \cdot \Size{\OPT(\sigma)} - C,
\end{equation} 

\vspace*{-0.5ex}\noindent%
for every possible input sequence $\sigma$, where $\Size{{\sf A}(\sigma)}$ is the number of bins covered %
by ${\sf A}$, $\OPT(\sigma)$ is a solution 
for which $\Size{\OPT(\sigma)}$ is maximal, and $C$ is some constant. 
It is useful to think of an online problem as a game played between the strategy and an {\em adversary}, that has full knowledge of the strategy and its decisions as the input items arrive. The adversary's objective is to produce the worst possible input sequence for the strategy. 

The {\em dual next fit strategy} (\DNF), maintains one active bin, into which it packs items until it is covered.  It then opens a new empty active bin 
and continues the process. For the one-dimensional case \DNF\ has a competitive ratio $1/2$, which is tight~\cite{ASSMANN1984:bincovering,CSIRIK1988163}.

However, if input items are bounded by some value $v_i \leq \alpha<1$, it is easy to show that \DNF\ has a competitive ratio as stated in the following inequality from Brodnik~{\em et~al}~\cite{matcos2022}. 
%

\vspace*{-2ex}
\begin{equation}\label{eq:dnf}
\Size{\DNF(\sigma_{\alpha})} > \frac{1}{1+\alpha}\Size{\OPTsa} - \frac{1}{1+\alpha}. 
\end{equation}

\vspace*{-0.5ex}\noindent
%
The strategy described in this paper assumes
the \emph{advice-on-tape} model~\cite{bockenhauer2017advice:advice}
in which
an \emph{oracle} has knowledge about both the strategy and the input sequence produced by the adversary, and writes a sequence of advice bits
on an unbounded \emph{advice tape} that the strategy can read at any time, before or while the requests are released by the adversary. The number of bits {\em read\/} by the strategy defines the \emph{advice complexity} of the strategy. Since the length of the advice is not explicitly given, the oracle needs some mechanism for the strategy to infer how many bits of advice it should read. This can be done with a self-delimiting encoding that extends the length of a bit string only by an additive lower order term~\cite[Section~3.2]{komm2016online}.

\paragraph*{Quantity Approximations using Fixed Point Numerical Values} 
Let $v$ have a binary fixed point representation
$v = \sum_{i=-\infty}^{q} v_i \cdot 2^i$,
where $q+1\geq1$ is the number of bits of the integer part of $v$ and $v_i$ is its $i^\text{th}$ bit.
We define $\dn{v}$ ($\up{v}$) to be the approximate value of $v$ obtained by taking the $b$ most significant bits of $v$ and replacing the remaining bits by zeroes (ones), 
i.e., $\dn{v}=\sum_{i=q-b+1}^{q}v_i \cdot 2^i$,
($\up{v}=2^{q-b+1}+\sum_{i=q-b+1}^{q}v_i \cdot 2^i$), 
with $b \geq 1$. Thus,

\vspace*{-1.5ex}
\begin{equation}\label{eqn:error}
v\! -\! \dn{v} = \left( \sum_{i=-\infty}^{q-b}v_i \cdot 2^i\right) \leq
           2^{-b+1} \cdot 2^{q}  
           \leq \tau v,
\end{equation}

\vspace*{-0.5ex}\noindent%
for some value $2^{-b+1}\leq\tau$. Hence, choosing 
$b \geq 
\log 1/\tau$, for sufficiently small $\tau$, guarantees that $\dn{v} \geq (1-\tau) v$.

Furthermore, $\dn{v}$ can be represented by $O(1/\tau +\log\log v)$ bits by providing the $b = 1/\tau$ most significant bits of $v$ and the length of the binary description of $v$, which has $O(\log\log v)$ size. The difference $\up{v}\!-\!v$ can be similarly bounded.

\section{Description of the Previous Strategies with Improvements}\label{sec:previous}

The Dual Harmonic strategy (\DHk) has been used as the building block for the advice based strategies previously published~\cite{boyar2021,matcos2022}. The strategy partitions the items by sizes into $k$ groups,
$]0,1/k[,~[1/k,1/(k-1)[,\ldots,[1/3,1/2[,~[1/2,1[$,
and packs items in each group according to~\DNF. The items of size $]0,1/k[$ are called {\em small items\/} and the items of size $[1/t,1/(t-1)[$ are called {\em$t$-items}, for integers $t\leq k$. Evidently, since \DHk\ is a pure strategy, it is at best $1/2$-competitive using the same argument as in Csirik and Totik~\cite{CSIRIK1988163}.

The partitioning of items into $k$ groups, as above, also facilitates in structuring the optimal solution. 
For a given input sequence $\sigma$ and an integer $k\geq2$, a fixed optimal covering \OPTs\ can be partitioned into groups, \Gttj, where the index $t_1 t_2\cdots t_j$, with $2\leq t_1\leq t_2\leq\cdots\leq t_j\leq k$, denotes that each bin in group \Gttj\ contains one $t_1$-item, one $t_2$-item, etc. Note that the multiplicity of an index denotes the number of times an item type occurs in the bin. If these items do not fill the bin, then it also contains the necessary amount of small items to do so. We further denote the group of bins that are only covered by small items by~\GS. For example, for $k=2$, the small items have size less than $1/2$ and $\OPTs=\Gtwotwo\cup\Gtwo\cup\GS$, while for $k=3$, the small items have size less than~$1/3$ and $\OPTs=\Gtwotwo\cup\Gtwo\cup\Gt{233}\cup\Gt{23}\cup\Gt{333}\cup\Gt{33}\cup\Gt{3}\cup\GS$.

Brodnik~{\em et~al.\@}~\cite{matcos2022} show that a modification of strategy \DHd\ to include advice gives an asymptotic competitive ratio of $2/3$. The idea of the strategy is that the oracle provides two values, $\big\lfloor|\Gtwo|/3\big\rfloor\leq m\leq\big\lfloor|\Gtwo|/2\big\rfloor$ and $x_m$, the size of the $m^{\rm th}$ largest item in the input sequence, to the strategy. The items of size at least $x_m$ are defined to be {\em good items}. The strategy then opens $m$ reserved bins, places 2-items of size at least $x_m$ into the reserved bins unless each reserved bin already contains a 2-item, and the remaining 2-items are packed two-by-two in separate bins. The small items are first used to pack the reserved bins up to a level of at least $1-x_m$ and once this is done, the remaining small items are packed using \DNF\ into separate bins. The remaining items are packed using \DNF\ into separate bins according to their group as specified by~\DHd.

The advice size for the strategy is $O(\zeta+\log n)$ bits of advice, where $\zeta$ is the number of bits required to represent the value $x_m$, they assume input items are rational. Since $m\leq n=|\sigma|$, $m$ can be represented with $O(\log n)$ bits and approximated with $O(\log\log n)$ bits \big(giving the competitive ratio $2/3-O(1/\log n)$\big). We remark that if the binary representation of $x_m$ includes a 0 early, say within the first $n$ or $\log n$ bits, then the oracle can truncate this binary representation a few bits further on and use a short representation as an approximation of $x_m$, giving that $\zeta\in O(\log n)$ or $O(\log\log n)$ depending on the case. This would then lead to the strategy having advice size $O(\log n)$ or $O(\log\log n)$.
Hence, for $\zeta$ to be very large, $x_m$ must be very close to~$1$.

\smallskip
Boyar~{\em et~al}.~\cite{boyar2021} consider a different computational model, where the input items are not necessarily assumed to be rational values of limited size. In fact, they use this assumption to obtain that $\Omega(n)$ bits of advice are necessary to achieve a competitive ratio above~$15/16$. They also modify \DHb\ to use $O(\log\log n)$ advice bits, considering only two item types, 2-items and small items, of size~$<1/2$, and achieve an asymptotic competitive ratio of~$8/15$. 

We give an overview of their strategy without providing a detailed analysis.
%
They define a covering to be {\em$(\alpha,\epsilon)$-desirable}, for $0<\alpha\leq1$ and an error limitation parameter $\epsilon$, if it obeys the following three properties:
\begin{description}
\item[Property~I.]
the covering has at least \aGt\ bins that each contain one 2-item and small items, 
\item[Property~II.]
the 2-items not placed according to Property~I are placed pairwise (possibly together with small items) in bins,
\item[Property~III.]
the small items not placed together with 2-items according to Property~II cover $2|\GS|/3-O(\epsilon|\Gtwo|)$ bins.
\end{description}

They prove that an $(\alpha,\epsilon)$-desirable covering, for any $\beta\geq1$, where $\Size{\Gtwotwo}+\Size{\Gtwo}=\beta\Size{\Gtwo}$, the number of covered bins is at least 
\begin{equation}\label{eqn:1d1CR}
\min\left\{\frac{\alpha+2\beta-1}{2\beta},\frac{2}{3}\right\}\cdot\big|\OPTs\big|
- O\big(\epsilon|\Gtwo|\big). 
\end{equation}
They claim a strategy such that if $\beta\geq15/14$, they can apply a lemma similar to Lemma~\ref{lem:dh2} below and claim a competitive ratio of~$8/15$.
If $\beta<15/14$ their main strategy yields an $(\alpha,\epsilon)$-desirable covering, for $\alpha=(7-6\beta)/15 - O(\epsilon)$ and $\epsilon>1/2^{b/2}>1/\log n$. To achieve this, they define $m=\big\lfloor|\Gtwo|/3\big\rfloor$ and the $\dn{m}$ largest items in the input sequence as {\em good items}. 
The subsequence of 2-items in the input is then split into three parts, the first two having sizes \dn{m} and the last part $2|\Gtwotwo|+|\Gtwo|-2\dn{m}$. The strategy obtains a $b$ bit approximation \dn{m} of the value $m$ from the oracle and opens \dn{m} reserved bins. The oracle additionally provides the strategy with information on which of the three subsequences to place in the reserved bins. If it is one of the first two, the \aGt\ largest 2-items are guaranteed to be good and the remaining ones are packed two-by-two using subsequent 2-items. If the oracle signals the last subsequence, slightly more than half of the 2-items in this sequence are placed in the reserved bins (since $\beta$ is small, these fit in the reserved bins). Again, the \aGt\ largest 2-items in the reserved bins are guaranteed to be good and the remaining 2-items are packed two-by-two using subsequent 2-items from the sequence. Analyzing the different cases 
gives that at least \aGt\ bins for $\alpha=(7-6\beta)/15 - O(\epsilon)$ among the reserved bins are covered with only one 2-item and small items. 

The small items are packed in the reserved bins using a {\em dual worst fit\/} (\DWF) strategy using no more than the total amount of small items that an optimal solution uses in the bins of \Gtwo\ that contain non-good 2-items. The particulars are similar to our presentation in Section~\ref{sec:smallitems} guaranteeing an $(\alpha,\epsilon)$-desirable packing. Setting $b=O(\log\log n)$, Expression~(\ref{eqn:1d1CR}) gives a competitive ratio of~$8/15-O(1/\log n)$ using $O(b)=O(\log\log n)$ bits of advice.

A first possible improvement to the Boyar~{\em et~al}.\@ strategy is to note that we can choose $m$ differently, as long as $2\dn{m}$ bins from \Gtwo\ 
contain enough small items to pack the $\dn{m}$ reserved bins to a sufficient level to guarantee that the resulting packing is $(\alpha,\epsilon)$-desirable, for some $\alpha$. The good items are hence defined to be the $|\Gtwo|-2\dn{m}$ largest ones in the input sequence. Analyzing this setup gives the optimal values of $m$ to be $m=\big\lfloor2|\Gtwo|/9+2|\Gtwotwo|/3\big\rfloor$ and $m=\big\lfloor|\Gtwo|/5+|\Gtwotwo|/2\big\rfloor$, both giving the competitive ratio~$5/9-O(1/\log n)$. The next section presents some additional modifications that improve the competitive ratio even further.


\section{Modification and Analysis of the Boyar~{\em et~al}.~Strategy}\label{sec:strategy}

In the remainder of this exposition, we improve on the analysis of the strategy by Boyar~{\em et~al}.~\cite{boyar2021} and show that with some fundamental changes, we significantly improve the competitive ratio, still using advice of the same order of magnitude. 

Let $T_2$ be the total number of 2-items in the input sequence $\sigma$, i.e., $T_2=2|\Gtwotwo|+|\Gtwo|$, since we, like Boyar~{\em et~al}~\cite{boyar2021}, only use two item types.
Further, define the parameter $\beta$ through $\Size{\Gtwotwo}+\Size{\Gtwo}=\beta\Size{\Gtwo}$, with $\beta \geq 1$. This leads to the following immediate result. 
\begin{lemma}{\rm(Corresponding to Lemma~1 in~\cite{boyar2021})}\label{lem:dh2}
When $|\Gtwo|=0$ or $\beta\geq121/107$, the strategy \DHb\ has competitive ratio at least~$135/242$.
\end{lemma}
\begin{proof}
Boyar~{\em et~al}.~\cite{boyar2021} prove that if $|\Gtwo|= 0$, then the strategy has competitive ratio $2/3$ and otherwise the competitive ratio of the strategy is $\min\big\{(2\beta-1)/(2\beta),2/3\big\}$. If $\beta\geq121/107$, the result follows.
\hfill$\Box$
\end{proof}
The oracle can signal with one bit to the strategy that $\beta\geq121/107$ and tell it to use the pure \DHb\ strategy guaranteeing a competitive ratio of at least~$135/242$. Therefore, we assume from now on that $|\Gtwo|>0$ and that~$\beta<121/107$. 

\smallskip
Let us define an {\em$(\alpha,\alpha',\rho,\epsilon)$-viable\/} covering for parameters $0\leq\alpha,\alpha'<1$, a 2-item size ratio $0\leq\rho\leq1$, and an error limitation parameter $\epsilon>0$ that will be assigned a value in Theorem~\ref{thm:main}, if the covering obeys the following three properties:
\begin{description}
\item[Property~I.]
the covering splits the subsequence of 2-items into two parts. The first $(1-\rho)T_2$ 2-items, such that the covering of these has at least \aGt\ bins, each containing one 2-item and small items, and the last subsquence of $\rho T_2$ 2-items, for which the covering has at least \apGt\ bins that each contain one 2-item and small items,
\item[Property~II.]
the 2-items from the first subsequence of $(1-\rho)T_2$ 2-items not placed according to Property~I are placed pairwise (possibly together with small items) in bins, (the last $\rho T_2$ 2-items are not placed pairwise in bins and these bins may therefore not necessarily be covered),
\item[Property~III.]
the small items not placed together with 2-items according to Properties~I and~II cover $2|\GS|/3-O\big(\epsilon|\Gtwo|\big)$ bins.
\end{description}

\noindent
We prove the following lemma.
\begin{lemma}\label{lem:viable}
For any input sequence $\sigma$ and any $\beta\geq1$, an $(\alpha,\alpha',\rho,\epsilon)$-viable covering, has at least the following number of covered bins
$$
\min\left\{\frac{(1-\rho)\big(2\beta-1\big)+\alpha+2\alpha'}{2\beta},\frac{2}{3}\right\}\cdot\big|\OPTs\big| - O\big(\epsilon|\Gtwo|\big).
$$
\end{lemma}
\begin{proof}
Let $a$ be the number of covered bins having exactly one 2-item from the first $(1-\rho)T_2$ 2-items in $\sigma$ and let $a'$ be the number of covered bins having exactly one 2-item from the last $\rho T_2$ 2-items. The first $(1-\rho)T_2$ are packed two-by-two except for the $a$ covered bins containing exactly one 2-item. Together with Property~III, we obtain
{\small
\begin{align*}
\frac{(1-\rho)T_2-a}{2}& + a + a' + \frac{2|\GS|}{3} - O\big(\epsilon|\Gtwo|\big)
\ = \
\frac{(1-\rho)T_2+a}{2} + a' + \frac{2|\GS|}{3} - O\big(\epsilon|\Gtwo|\big)
\\
&\geq
\frac{(1-\rho)\big(2|\Gtwotwo|+|\Gtwo|\big)+\alpha|\Gtwo|}{2} + \alpha'|\Gtwo| +  \frac{2|\GS|}{3}-O\big(\epsilon|\Gtwo|\big)
\\
&=
\frac{(1-\rho)\big(2\beta-1\big)+\alpha+2\alpha'}{2}|\Gtwo| + \frac{2}{3}|\GS|-O\big(\epsilon|\Gtwo|\big)
\\
&\geq
\min\left\{\frac{(1-\rho)\big(2\beta-1\big)+\alpha+2\alpha'}{2\beta},\frac{2}{3}\right\}\cdot\big|\OPTs\big| - O\big(\epsilon|\Gtwo|\big)
\end{align*}
}%
bins, since, for any pair of non-negative values, $I$ and $J$, $I\!+\!J\!\geq\!\min\{I,J\}$ and $|\OPTs|\!=\!|\Gtwotwo|+|\Gtwo|+|\GS|=\beta|\Gtwo|+|\GS|$.
\hfill$\Box$
\end{proof}
\noindent
We denote the modified strategy that we present by~\DHbb, where $b$ is the number of bits used to approximate advice values. We fix $\epsilon>1/2^{b/2}$, use $b$ bit fixed point approximations as described in Section~\ref{sec:prel} and will specify the value of $b$ later. 

Let $\mR\defeq\Big\lfloor(1-\epsilon)^2\big(27|\Gtwo|/121+2|\Gtwotwo|/3\big)\Big\rfloor$ and $m\defeq\big\lceil(1+\epsilon)\mR\big\rceil$. Define $\nG\defeq|\Gtwo|-2m$ and let the \nG\ largest items of the input sequence $\sigma$ be the {\em good 2-items}. The oracle transmits the advice value \dn\mR\ and the strategy opens \dn\mR\ bins called {\em reserved bins}. It repeats the following steps for each item~$v$.
\begin{description}
\item[{\bf if} $v$ {\bf is small}] place it according to the presentation in Section~\ref{sec:smallitems},
\item[{\bf else}]
($v$ is a 2-item), either place it in one of the reserved bins that does not already contain a 2-item according to the presentation in Section~\ref{sec:2items}, ensuring that at least \aGt\ of them are good 2-items; or place it in a reserved bin that contains a 2-item that is not good; or place it pairwise with another 2-item in a new bin.
\end{description}
%
%

\subsection{Placing the 2-items}\label{sec:2items}
We show the following result.
\begin{lemma}\label{lem:2-items} 
For $\beta < 121/107$, there exist threshold values $\delta^T>0$,
$\alpha_L^T>0$, and size ratios $\rho>0$ and $\rho'>0$, such that the strategy \DHbb\ using $O(b)$ bits of advice produces an
\begin{enumerate}\vspace*{-2ex}
\item
$(\alpha,\epsilon)$-desirable covering, if $\delta\leq\delta^T$,
\item
$(\alpha_L^T,\delta,\rho,\epsilon)$-viable covering, if $\delta>\delta^T$ and $\alpha_L\geq\alpha_L^T$,
\item
$(0,\alpha\!-\!\alpha_L^T\!+\!\delta,\rho',\epsilon)$-viable covering, if $\delta>\delta^T$ and $\alpha_L<\alpha_L^T$,
\end{enumerate}\vspace*{-1ex}
when $\alpha\leq685/1452-\beta/3-\delta/4-\epsilon/4$, for any $\epsilon>1/2^{b/2}$\!.
\end{lemma}
\begin{proof}
We only prove the appropriate properties~I and~II for the three cases stated in the Lemma and defer the proof of Property~III to Lemma~\ref{lem:1Dsmall} in Section~\ref{sec:smallitems}.

We implicitly subdivide the sequence of 2-items into four consecutive subsequences. The first three subsequences contain exactly \dn{\mR}\ 2-items each and the last subsequence consists of the remaining 2-items. The oracle can signal using no more than two bits of advice which of the four subsequences the strategy should use to place in the reserved bins. If the strategy chooses the last subsequence the oracle can, with two additional bits of advice, identify whether the strategy should produce an $(\alpha,\epsilon)$-desirable covering or one of two possible $(\alpha,\alpha',\rho,\epsilon)$-viable coverings. We have four cases.

\smallskip
\CASE{Case~1}
If the oracle signals that one of the first three subsequences contains at least $\dn\aGt-1$ good 2-items, using two bits and the approximate value $\dn\aGt$, then the strategy places the 2-items in previous subsequences two-by-two in separate bins, thus guaranteeing that these are covered. The 2-items that the oracle designates are placed in the reserved bins and once this is done, the strategy declares the largest $\dn\aGt-1$ of them to be good. It then packs the other non-good 2-items in the reserved bins with subsequent 2-items, and 
places any further 2-items in the input sequence two-by-two in separate bins. This gives an $(\alpha,\epsilon)$-desirable covering.

\smallskip
\CASE{Case~2}
If the oracle signals the last subsequence, then each of the previous subsequences contains fewer than $\dn\aGt-1$ good 2-items and the strategy places the 2-items in these  subsequences two-by-two in separate bins. There are $Z=T_2-3\dn\mR$ 2-items left in the last subsequence, which in turn is further partitioned into three {\em chunks}, that consist of $\dn{X_L}$, $\dn{X_R}$, and $Y$ 2-items respectively. 
The relationships between these values are: $\dn{X_L}+\dn{X_R}+Y=Z$, $\dn{X_L}+\dn{X_R}-Y\leq\dn\aGt$\!, $\dn{X_L}+\dn{X_R}\leq\dn{\mR}$, $\dn{X_L}+Y\leq\dn{\mR}$, and $\dn{X_R}+Y\leq\dn{\mR}$. The initial two chunks contain at least \dn\aGt\ good items, separated into \dn\aLGt\ good items in the first chunk and at least $\dn\aGt\!\!-\!\dn\aLGt$ in the second chunk. The third chunk contains \dGt\ good 2-items.

Let $\delta^T$ be a threshold value to be determined later. If $\delta\leq\delta^T$, the oracle signals the strategy to perform the steps of the next case.

\smallskip
\CASE{Case~2a}
The oracle transmits the values \dn{X_L}, \dn{X_R}, and the value \dn\aGt\ to the strategy on the advice tape. The strategy places the first $\dn{X_L}+\dn{X_R}$ 2-items (all the items in the first two chunks) in the reserved bins, and declares the largest $\dn\aGt\!\!-\!1$ of them to be good. Use the remaining $Y$ 2-items in the sequence (the last chunk) to pack together with the 2-items in the reserved bins that were not declared good.
The 2-items in the first two chunks fit in the reserved bins since $\dn{X_L}+\dn{X_R}\leq\dn{\mR}$ and we show that they contain at least $\dn\aGt\!\!-\!1$ good 2-items. 

Let \gta\ be the number of good 2-items placed alone 
in the reserved bins. The final $Y$ 2-items in the input sequence contain $\dGt$ good items, hence

\vspace*{-3ex}
{\small
\begin{align*}
\gta
&\geq
\nG - 3\big(\dn\aGt-2\big)-\dGt
\ \ \geq \ \
|\Gtwo| - 2m - 3\cdot\aGt - \dGt + 6
\\
&\geq
|\Gtwo| - 2(1+\epsilon)\mR - 3\cdot\alpha\cdot|\Gtwo| - \delta\cdot|\Gtwo| + 2
\\
&\geq
|\Gtwo| - 2(1+\epsilon)
\Big\lfloor(1-\epsilon)^2\big(27|\Gtwo|/121+2|\Gtwotwo|/3\big)\Big\rfloor
- 3\alpha\cdot|\Gtwo| - \delta\cdot|\Gtwo| + 2
\\
&\geq
67|\Gtwo|/121 - 4(\beta-1)\cdot|\Gtwo|/3 
- 3\alpha\cdot|\Gtwo| - \delta\cdot|\Gtwo| - \epsilon\cdot|\Gtwo| - 1
\\
&\geq
\alpha\cdot|\Gtwo| - 1
\ \ \geq \ \
\aGt - 1
\ \ \geq \ \
\dn\aGt - 1,
\end{align*}
}%

\vspace*{-0ex}\noindent
if $\alpha\leq685/1452-\beta/3-\delta/4-\epsilon/4$. Thus, we obtain an $(\alpha,\epsilon)$-desirable covering.
%


\smallskip
\CASE{Case~2b}
If $\delta>\delta^T$, we consider the threshold value $\alpha_L^T$, also to be determined later. If $\alpha_L\geq\alpha_L^T$, the oracle signals this fact to the strategy in addition to the values \dn{X_L} and \dn{\big\lfloor\alpha_L^T\!\cdot\!|\Gtwo|\big\rfloor}\!. The strategy places the 2-items in the first chunk in the reserved bins and declares $\dn{\big\lfloor\alpha_L^T\!\cdot\!|\Gtwo|\big\rfloor}\!$ of them to be good. It uses the 2-items in the second chunk to pack the declared non-good 2-items in the reserved bins two-by-two and packs the remaining 2-items in the second chunk two-by-two into separate bins. The 2-items of the third chunk are then placed in the remaining reserved bins. Thus, the strategy produces a $(\alpha_L^T,\delta,\rho,\epsilon)$-viable covering, with $\rho=Y/T_2$.

\smallskip
\CASE{Case~2c}
If $\delta>\delta^T$ and $\alpha_L<\alpha_L^T$, then the oracle signals this fact to the strategy in addition to the value \dn{X_L}. 
The strategy packs 2-items in the first chunk of the last group two-by-two in separate (not the reserved) bins and places the 2-items of the last two chunks in the reserved bins. Hence, the strategy gives an $(0,\alpha_R\!+\!\delta,\rho',\epsilon)$-viable covering, with $\rho'=(\dn{X_R}\!+\!Y)/T_2$.
\hfill$\Box$
\end{proof}

\subsection{Placing the Small Items}\label{sec:smallitems}


As before, we let $\epsilon>1/2^{b/2}$\!, with $b$ being the number of bits used to approximate integers given as advice, and let $\mR\defeq\Big\lfloor(1-\epsilon)^2\big(27|\Gtwo|/121+2|\Gtwotwo|/3\big)\Big\rfloor$ be the number of reserved bins that the oracle proposes to the strategy. The strategy hence transmits the approximate value \dn\mR\ to~\DHbb. Next, let $m\defeq\big\lceil(1+\epsilon)\mR\big\rceil$ and let $1-d$ be the size of the 
$(|\Gtwo|-2m)^{\rm th}$ 
input item in descending sorted order of size, i.e., the size of the smallest good item. By a simple exchange argument, we may assume that the bins in \Gtwo\ contain the $|\Gtwo|$ largest 2-items in the input sequence~$\sigma$, since any pair of 2-items will cover a bin in~\Gtwotwo.

Using the terminology in~\cite{boyar2021} we define small items of size at least $\up d$ as {\em black items}, and those of size less than $\up d$ as {\em white items}. Now, let \nB\ be the number of bins in \Gtwo\ (of the fixed optimal solution) that contain black items.

\subsubsection{Placing Black Items}

Since \Gtwo\ contains \nB\ bins having black items it is possible to cover $\mRB\defeq\min\{\dn\mR,\nB\}$ reserved bins with exactly one good item (of size $\geq1-d$) and one black item (of size $\geq\up d$). Our objective is therefore to fill these reserved bins with the smallest black items. Again by an exchange argument, we can choose the smallest \nB\ such items to be in bins of \Gtwo\ and the remaining ones to be in bins of \GS\ without decreasing the size of the solution. 

Let \sB\ be the size of the ${\mRB}^{\rm th}$ smallest black item in the input sequence, we let \xB\ be the number of items among the \mRB\ smallest black items that have size at most $\dn\sB$ and let \eB\ be the number of items among the \mRB\ smallest black items that have size in the range $]\dn\sB,\up\sB]$, so $\mRB=\xB+\eB$. Let $\mB\defeq\xB+\dn\eB$, following the presentation in~\cite{boyar2021}.

The \DHbb\ strategy receives the advice \up d\!, $\dn\mB$\!, $\dn\sB$\!, and $\dn\eB$ from the oracle to deal with the black items.
Out of the \dn\mR\ reserved bins, the strategy allocates \dn\mB\ of them as {\em black reserved bins}, each containing one black item. When a black item arrives of size at most $\dn\sB$ it is placed in a black reserved bin. When a black item in the range $]\dn\sB,\up\sB]$ arrives it is also placed in a black reserved bin unless \dn\eB\ such items have already been placed in the black reserved bins. 
We have two cases to take care of. 

If $\sB\geq\epsilon$, then $\up\sB<\sB+1/2^b\leq\sB+\epsilon^2\leq(1+\epsilon)\sB$. 
Hence, at most \dn\eB\ black items are used that have size $\up\sB\leq(1+\epsilon)\sB$, i.e., a factor of $(1+\epsilon)$ larger than the ${\mRB}^{\rm th}$ smallest black item.

If $\sB<\epsilon$, on the other hand, then the strategy packs a total amount of at most $\up\sB(\xB+\dn\eB)<(\epsilon+\epsilon^2)\mRB\leq2\epsilon\cdot\mRB\leq2\epsilon\cdot|\Gtwo|$ black items.

Once the black reserved bins have been filled with a black item, the remaining black items that arrive in the input sequence are placed together with other small items using \DNF\ into separate bins.

\subsubsection{Placing White Items}

Let $\mW=\dn\mR-\dn\mB$ be the number of remaining reserved bins, we call them {\em white reserved bins}. The strategy places white items as they arrive into these bins using the Dual Worst Fit strategy (\DWF) until each remaining white reserved bin contains white items to a level of at least $\up d$\!. When a white item arrives, the substrategy \DWF\ places the item in the white reserved bin that contains the smallest sum of white items. Once each white reserved bin contains white items to the amount of $\up d$\!, the remaining white items that arrive in the input sequence are placed together with other small items using \DNF\ into separate bins as in the case with the black items.


We prove the following results as in~\cite{boyar2021}.

\begin{lemma}\label{lem:white} {\rm(Corresponding to Lemma~3 in~\cite{boyar2021})}
A reserved bin that includes a good item will be covered when \DHbb\ terminates.
\end{lemma}

\vspace*{-2ex}
\begin{proof}
By definition, a good item has size at least $1-d$ so it is sufficient to show that each reserved bin contains small items to the amount of at least $d$. This is clearly true for the black reserved bins, since $\up d\geq d$. 

Consider the bins in \Gtwo\ of the optimal solution. Of these, $|\Gtwo|-\nG=2m$ bins contain a 2-item of size at most $1-d$ and at least $2m-\mRB$ contain only white small items. The bins in \Gtwo\ therefore contain white items to the amount of $S_W\geq d(2m-\mRB)$. 

The \DWF\ strategy stops placing white items when each white reserved bin contains at least $\up d$ amount of white items or there are no more white items in the input sequence. 
Assume for a contradiction that all the white items in the input sequence are placed in the white reserved bins, but that not all the white reserved bins contain at least $d$ amount of white items, i.e., the input sequence reaches the end without having placed at least $d$ amount of white items in each of the white reserved bins.

Let $a_1,\ldots,a_l$ be the sizes of the white items that have size at least $d$. Note that $l<\mW$, otherwise all white reserved bins have at least $d$ amount of white items, a contradiction. The remaining white items have size less than $d$. The white reserved bin that contains the item of size $a_i$ when \DWF\ has processed all the white items contains less than $d+a_i$ amount of white items since \DWF\ always places white items in the bin containing the least amount of white items, for $1\leq i\leq l$. For the same reason, any other white reserved bin contains less than $2d$ amount of white items.
This gives us the following bound on~$S_W$\!,

\vspace*{-1.5ex}
{\small
\begin{equation*}
S_W
<
\sum_{i=1}^l(a_i + d) + \sum_{i=l+1}^{\mW}2d
=
2\mW\cdot d + \sum_{i=1}^l (a_i-d).
\end{equation*}
}

\vspace*{-0ex}\noindent
On the other hand, since those bins in \Gtwo\ that do not contain good items must each have at least $d$ amount of small items in them. Discounting those bins that contain black items, there are $2m-\mRB\geq2\mW$ such bins in~\Gtwo, giving us a lower bound on the amount of white items in bins of~\Gtwo. We obtain

\vspace*{-1.5ex}
{\small
\begin{equation*}
S_W
\geq
d(2\mW-l) + \sum_{i=1}^l a_i
=
2\mW\cdot d + \sum_{i=1}^l (a_i-d),
\end{equation*}
}

\vspace*{-0ex}\noindent
giving us a contradiction. Hence, every white reserved bin contains white items to the amount of at least $d$ if the \DWF\ strategy reaches the end of the input sequence before terminating.
Thus, the \DWF\ strategy guarantees that a white reserved bin contains at least $d$ and less than $2\up d$ amount of white items. We have, if $d\geq\epsilon$, that

\vspace*{-3ex}
{\small
\begin{align*}
2\up d\mW
&
=
2\up d\big(\dn\mR - \dn\mB\big)
\ \
\leq
\ \
2\up d\big(\dn\mR - (1-\epsilon)\mB\big)
\\
&
=
2\up d\big(\dn\mR - (1-\epsilon)(\xB+\dn\eB)\big)
\!\!\ \
\leq
\ \
\!2\up d\big(\dn\mR - (1-\epsilon)^2(\xB+\eB)\big)
\\
&
\leq
2\up d\big(\mR - (1-\epsilon)^2\mRB\big)
\ \
\leq
\ \
2\big(d+1/2^b\big)\big(\mR - (1-\epsilon)^2\mRB\big)
\\
&
<
2\big(d+\epsilon^2\big)\big(\mR - (1-\epsilon)^2\mRB\big)
\ \
\leq
\ \
2d\big(1+\epsilon\big)\big(\mR - (1-\epsilon)^2\mRB\big)
\\
&
\leq
d\big(2(1+\epsilon)\mR  - 2(1+\epsilon)(1-\epsilon)^2\mRB\big)
\ \
\leq
\ \
d\big(2m - \mRB\big) \ \ \leq \ \ S_W,
\end{align*}
}%

\vspace*{-1ex}\noindent%
which holds for $1/2^{b/2}<\epsilon$ sufficiently small. Hence, there is sufficient amount of white items among those in bins of \Gtwo\ to place among the reserved white bins and reach $\up d$ white items in each such bin, thus guaranteeing that the \DWF\ strategy terminates in this case.

If $d<\epsilon$, then following the same reasoning as above, we have

\vspace*{-3ex}
{\small
\begin{align*}
2\up d\mW
&
\leq \
2(d+1/2^b)\mR 
\ \
<
\ \
2(d+\epsilon^2)\mR 
\ \
<
\ \
2(\epsilon+\epsilon^2)\mR
\ \
\leq
\ \
3\epsilon\mR
\\
&
\leq
\ \
3\epsilon(1-\epsilon)^2\big(27|\Gtwo|/121 + 2|\Gtwotwo|/3\big)
< \
12055\cdot \epsilon|\Gtwo|/ 12947
\ \ \leq \ \ 
\epsilon\cdot|\Gtwo|,
\end{align*}
}%

\vspace*{-1ex}\noindent%
since $\beta<121/107$ and $1/2^{b/2}<\epsilon$ sufficiently small, concluding the proof.
\hfill$\Box$
\end{proof}

\begin{lemma}\label{lem:1Dsmall} {\rm(Corresponding to Lemma~4 in~\cite{boyar2021})}
The strategy \DHbb\ covers at least $2|\GS|/3-O(\epsilon|\Gtwo|)$ small bins. 
\end{lemma}
\begin{proof}
From the proof of the previous lemma, if $d\geq\epsilon$, we can view all the white items placed in the reserved bins by the strategy to come from bins in \Gtwo\ in the optimal solution and, if $d<\epsilon$, that the sum of all the white items in bins of \Gtwo\ is at most~$\epsilon\cdot|\Gtwo|$. 
In either case, the bins in \GS\ have not been reduced by more than $\epsilon\cdot|\Gtwo|$ amount of small items to place in the white reserved bins.

As for the black items, the strategy uses \mRB\ of the smallest black items to place in the black reserved bins, except possibly in \dn\eB\ cases when such an item lies in the range $]\dn\sB,\up\sB]$, where the correct black item would have size just above \dn\sB\ but the strategy chose an item of size no more than \up\sB. The amount of small items we overuse to fill the black reserved bins is at most
\begin{equation*}
1/2^b\cdot\dn\eB<\epsilon^2\cdot\dn\eB\leq\epsilon\cdot\mRB\leq\epsilon\cdot|\Gtwo|.
\end{equation*}
Again, the bins in \GS\ have not been reduced by more than $\epsilon\cdot|\Gtwo|$ amount of small items to place in the black reserved bins.

Hence, the amount of small items that remains for the strategy to use is at least $|\GS|-2\epsilon\cdot|\Gtwo|$
and by Inequality~(\ref{eq:dnf}), these small items cover at least $2|\GS|/3-O(\epsilon|\Gtwo|)$ bins, proving the lemma.
\hfill$\Box$
\end{proof}


\noindent
We can now prove our main result.
\begin{theorem}\label{thm:main}
The modified strategy\/ \DHllb\!, using $O(\log\log n)$ bits of advice, 
achieves an asymptotic competitive ratio of at~least $135/242-O(1/\log n)\approx0.5578\ldots-O(1/\log n)$.
\end{theorem}
\begin{proof}
If $\beta\geq121/107$ the result follows from Lemma~\ref{lem:dh2} and the oracle can signal this case using one bit. If $1\leq\beta<121/107$, the oracle provides a constant number of integer values, all smaller than $n$, and the threshold values \up{d} and $\dn\sB$\!. In addition, the oracle provides a constant number of selection bits of advice to help the strategy decide between the appropriate strategy cases.

From Lemma~\ref{lem:2-items}, we know that there exist parameters $\delta^T$, $\alpha_L^T$, $\rho$, and $\rho'$ such that the strategy, with appropriate advice will produce desirable or viable coverings as claimed, for $\alpha$ sufficiently small. Let $\delta^T=1/11$, $\alpha_L^T=5\alpha/14$, $\rho=13/121$ and $\rho'=26/121$, chosen to balance the cases in Lemma~\ref{lem:2-items}.

If $\delta\leq\delta^T=1/11$, then the strategy produces an $(\alpha,\epsilon)$-desirable covering in Cases~1 and~2a of the proof of Lemma~\ref{lem:2-items}. Choosing $\alpha=685/1452-\beta/3-\delta/4-\epsilon/4$, as in the prerequisite of the lemma, Expression~(\ref{eqn:1d1CR}) gives a competitive ratio of
{\small
\begin{equation*}
\frac{\alpha+2\beta-1}{2\beta} 
=
\frac{\dfrac{685}{1452}-\dfrac{\beta}{3}-\dfrac{\delta}{4}-\dfrac{\epsilon}{4}+2\beta-1}{2\beta}
\geq
\dfrac{5}{6} - \dfrac{100}{363\beta} - \epsilon
\geq 
\frac{135}{242} - O\!\left(\!\frac{1}{\log n}\!\right).
\end{equation*}%
}%

If $\delta>\delta^T=1/11$ and $\alpha_L\geq\alpha_L^T=5\alpha/14$, then the strategy produces an $(\alpha_L^T,\delta,\rho,\epsilon)$-viable covering in Case~2b of the proof of Lemma~\ref{lem:2-items}. With $\alpha_L^T=5\alpha/14=3425/20328-5\beta/42-5\delta/56-5\epsilon/56$, Lemma~\ref{lem:viable} gives a competitive ratio of
{\small
\begin{equation*}
\frac{(1-\rho)\big(2\beta-1\big)+\alpha_L^T+2\delta}{2\beta}
\geq
\frac{8467}{10164} - \frac{2797}{10164\beta} - \epsilon
\geq
\frac{135}{242} - O\!\left(\!\frac{1}{\log n}\!\right).
\end{equation*}%
}%

If $\delta>\delta^T=1/11$ and $\alpha_L<\alpha_L^T=5\alpha/14$, then the strategy produces an $(0,\alpha_R+\delta,\rho',\epsilon)$-viable covering in Case~2c of the proof of Lemma~\ref{lem:2-items}. Again, with $\alpha_L^T=5\alpha/14$, Lemma~\ref{lem:viable} gives a competitive ratio of
{\small
\begin{align*}
\frac{(1-\rho')\big(2\beta-1\big)+2(\alpha_R+\delta)}{2\beta}
\geq
\frac{(1-\rho')\big(2\beta-1\big)+2(\alpha-\alpha_L^T+\delta)}{2\beta} \geq
\hspace*{-40.75ex}&
\\
&
\geq
\frac{967}{1694} - \frac{1}{77\beta} - \epsilon
\ \geq \
\frac{135}{242} - O\!\left(\!\frac{1}{\log n}\!\right).
\end{align*}%
}%

By setting $b=2\log\log n$ and $\epsilon=2/2^{b/2}=2/\log n$, the strategy can approximate any integer of size up to $n=|\sigma|$, proving the claim.
\hfill$\Box$
\end{proof}

\section{Conclusions}

We provide a deeper analysis and improvements to the strategy for online bin covering by Boyar~{\em et~al}.~\cite{boyar2021} that uses $O(\log\log n)$ bits of advice and achieves a significantly better competitive ratio of $135/242-O(1/\log n)$, where $n$ is the length of the input sequence. 

Improving the competitive ratio further using $O(\log\log n)$ advice bits is an interesting challenge. 

The authors would like to thank the anonymous reviewers for their comments and constructive criticism of an earlier draft of this manuscript.

{
}


%
%
%
%
%
%
%

\end{document}